\documentclass[aps,prl,floatfix,twocolumn,showpacs]{revtex4}%
\usepackage{amsfonts}
\usepackage{amsmath}
\usepackage{graphicx}
\usepackage{amssymb}%
\begin{document}
\title{Superconducting proximity effect in a diffusive ferromagnet with spin-active interfaces}
\author{A. Cottet and W. Belzig}
\affiliation{Department of Physics and Astronomy, University of Basel, Klingelbergstrasse
82, 4056 Basel, Switzerland}

\begin{abstract}
We reconsider the problem of the superconducting proximity effect in a
diffusive ferromagnet bounded by tunneling interfaces, using spin-dependent
boundary conditions. This introduces for each interface a phase-shifting
conductance $G_{\phi}$ which results from the spin dependence of the phase
shifts acquired by electrons upon scattering on the interface. We show that
$G_{\phi}$ strongly affects the density of states and supercurrents predicted
for superconducting/ferromagnetic hybrid circuits. We show the relevance of
this effect by identifying clear signatures of $G_{\phi}$ in the data of
T.~Kontos \textit{et al} [Phys. Rev. Lett. \textbf{86}, 304 (2001),
\textit{ibid.} \textbf{89}, 137007 (2002)].

\end{abstract}
\date{\today}
\maketitle


Superconducting/ferromagnetic (S/F) hybrid structures raise the fundamental
question of what happens when two phases with different broken symmetries
interact. When a F metal with uniform magnetization is connected to a BCS
superconductor, the singlet electronic correlations characteristic of the S
phase propagate into F via Andreev reflections which couple electrons and
holes with opposite spins and excitation energies. In the diffusive case, this
propagation occurs on a scale limited by the ferromagnetic exchange field. The
decay of the correlations in F is accompanied by oscillations of the
superconducting order parameter because the exchange field induces an energy
shift between the correlated electrons and holes \cite{Buzdin1982,Golubov}.
This has been observed experimentally through oscillations of the density of
states (DOS) in F \cite{TakisN}, or of the critical current $I_{0}$ through
S/F/S structures \cite{Ryazanov,TakisI,SellierPRB,Blum}, with the thickness of
F or the temperature. Remarkably, the oscillations of $I_{0}$ have allowed to
obtain $\pi$-junctions, i.e. Josephson junctions with $I_{0}<0$
\cite{Guichard}, which could find applications in the field of superconducting
circuits \cite{Ioffe}.

The interface between a ferromagnet and a non-magnetic material can scatter
electrons with spin parallel or antiparallel to the magnetization of the
ferromagnet with different phase shifts. The \textit{spin-dependence} of the
interfacial phase shifts (SDIPS) is a general concept in the field of
spin-dependent transport.The SDIPS implies that spins non collinear to the
magnetization precess during the scattering by the interface. This so-called
spin mixing is expected to affect drastically the behavior of F/normal metal
\cite{FNF} when several F electrodes with non collinear magnetization are
used, as observed experimentally by \cite{Pratt}. The same phenomenon is
predicted to occur in F/coulomb blockade island \cite{CB}, and F/Luttinger
liquid \cite{Luttinger} hybrid circuits. In S/F hybrid systems
\cite{Tokuyasu,otherBC,demoBC}, the SDIPS is even predicted to affect the
system in collinear configurations, due to the coupling of electrons and holes
with opposite spins by the Andreev reflections. However, few experimental
signatures of the SDIPS have been identified up to now (e.g. Ref.
\cite{Tokuyasu} proposes for the data of \cite{Tedrow} an interpretation based
on the\ SDIPS).

In this Letter, we reconsider the problem of the superconducting proximity
effect in a diffusive F. Up to now the tunnel S/F contacts used to produce
this effect were described (see e.g. \cite{Golubov}) with spin independent
boundary conditions (BC) derived in \cite{Kuprianov} for the spin-degenerate
case. Instead of that, we use spin-dependent BC based on Ref. \cite{demoBC}.
These BC introduce a phase shifting conductance $G_{\phi}$ which takes into
account the SDIPS. We show that $G_{\phi}$ strongly affects the phase and the
amplitude of the oscillations of the DOS or $I_{0}$ with the thickness of F.
Our approach thus provides a framework for future work on S/F diffusive
circuits with tunneling interfaces. We show its relevance by a comparison with
the data of \cite{TakisN,TakisI} which shows that strong experimental
manifestations of the SDIPS have already been observed through the
superconducting proximity effect.

We consider a S/F hybrid circuit with a single F electrode homogeneously
magnetized in direction $\overrightarrow{z}$. In the diffusive limit, the
electrons in a superconducting or ferromagnetic electrode $\alpha$ can be
described with quasi-classical and diffusive Green's functions $\check
{G}_{\alpha}$ in the Keldysh$\otimes$Nambu$\otimes$spin space (we use the
notations of \cite{demoBC}). The BC at a S/F interface can be calculated by
assuming that the interface potential locally dominates the Hamiltonian, i.e.
at a short distance it causes only ordinary scattering (with no particle-hole
mixing). We characterize this scattering with transmission and reflection
amplitudes $t_{n,\sigma}^{S(F)}$ and $r_{n,\sigma}^{S(F)}$ for electrons
coming from the S(F) side in channel $n$ with a spin $\sigma$ parallel
($\sigma=\uparrow$) or antiparallel ($\sigma=\downarrow$) to $\overrightarrow
{z}$. In practice, the planar S/F contacts used to induce the superconducting
proximity effect in a diffusive ferromagnet are likely to be in the tunnel
limit \cite{Geers,Takis2004}, due e.g. to a mismatch of band structure between
S and F, thus we assume $T_{n}=\sum_{\sigma}|t_{n,\sigma}^{S}|^{2}\ll1$. We
also consider that the system is weakly polarized. Following
\cite{demoBC,CircuitTheory}, the BC at the right hand side F of a S/F
interface is
\begin{align}
2g_{F}\check{G}_{F}\frac{\partial\check{G}_{F}}{\partial x} &  =\left[
G_{t}\check{G}_{S}+iG_{\phi}\sigma_{z}\check{\tau}_{3}+\frac{G_{MR}}{2}%
\check{D}_{+},\check{G}_{F}\right]  \nonumber\\
&  +\left[  iG_{\chi}\check{G}_{S}\check{D}_{-}+iG_{\xi}\check{D}_{-}\check
{G}_{F},\check{G}_{F}\right]  \label{bc}%
\end{align}
with $\check{D}_{\pm}=\sigma_{z}\check{\tau}_{3}\check{G}_{S}\pm\check{G}%
_{S}\sigma_{z}\check{\tau}_{3}$. Here, $\sigma_{z}$ and $\check{\tau}_{3}$ are
Pauli matrices in spin and Nambu space respectively. The conductivity of $F$
times area of the junction, noted $g_{F}$, is assumed to be spin independent.
The coefficient $G_{t}=G_{Q}\sum\nolimits_{n}T_{n}$ is the tunneling
conductance, $G_{MR}=G_{Q}\sum\nolimits_{n}(|t_{n,\uparrow}^{S}|^{2}%
-|t_{n,\downarrow}^{S}|^{2})$ is the magnetoresistance term which leads to a
spin-polarization of the current, and $G_{\phi}=2G_{Q}\sum\nolimits_{n}%
(\rho_{n}^{F}-4[\tau_{n}^{S}/T_{n}])$ is the phase-shifting conductance, with
$\tau_{n}^{S}=\operatorname{Im}[t_{n,\uparrow}^{S}t_{n,\downarrow}^{S~\ast}]$,
$\rho_{n}^{F}=\operatorname{Im}[r_{n,\uparrow}^{F}r_{n,\downarrow}^{F~\ast}]$
and $G_{Q}=e^{2}/h$. These three terms already appeared in \cite{demoBC} for
studying normal electrodes in contact with S and F reservoirs (with no
proximity effect in F). The extra terms in $G_{\xi}=-G_{Q}\sum\nolimits_{n}%
\tau_{n}^{S}$ and $G_{\chi}=G_{Q}\sum\nolimits_{n}T_{n}(\rho_{n}^{F}+\tau
_{n}^{S})/4$ occur because there are superconducting correlations at both
sides of the interface. Note that $G_{\phi}$, $G_{\chi}$ and $G_{\xi}$ can be
finite only if the phase shifts acquired by the electrons upon reflection or
transmission at the interface are spin-dependent. \begin{figure}[ptb]
\includegraphics[width=0.93\linewidth]{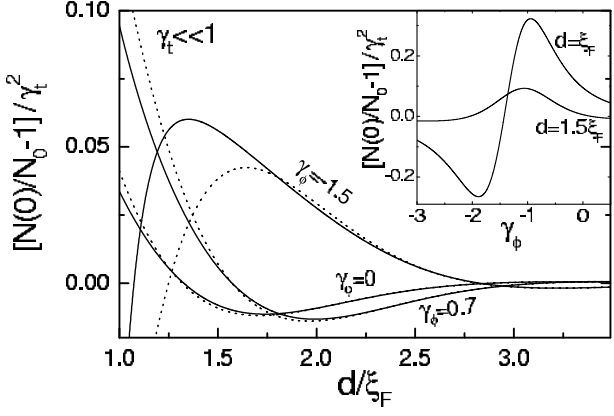}\caption{Zero energy density
of states at $x=d$ in a S/F/I structure, in terms of $[(N(0)/N_{0}%
)-1]/\gamma_{t}^{2}$ as a function of $d/\xi_{F}$, for $\gamma_{t}\ll1$ and
different values of $\gamma_{\phi}$ (full lines). The dotted lines show
$4[(N(0)/N_{0})-1]/\gamma_{t}^{2}$ at $x=d$ in a semi-infinite S/F structure
with the same values of $\gamma_{t}$ and $\gamma_{\phi}$. The inset shows the
DOS at $x=d$ as a function of $\gamma_{\phi}$ for the S/F/I structure.}%
\label{Graph1}%
\end{figure}The exact values of these conductance coefficients depend on the
microscopic structure of the interface. However, we can estimate their
relative orders of magnitude in a rectangular potential barrier model, by
describing the ferromagnetism of F with an exchange field $E_{ex}$ much
smaller than the spin averaged Fermi energy $E_{F}$ of F. This gives
expressions of $G_{MR}$, $G_{\phi}^{\alpha}$, $G_{\chi}^{\alpha}$ and $G_{\xi
}^{\alpha}$ linear with $E_{ex}/E_{F}$. The tunnel limit can be reached by
considering a strong mismatch between the Fermi wavevectors in S and F (case
1) or a high enough barrier (case 2). In both limits we find $\left\vert
G_{MR}\right\vert ,\left\vert G_{\chi}\right\vert ,\left\vert G_{\xi
}\right\vert \ll G_{t}$, which allows us to neglect these terms in the
following. In case 1, we find $\left\vert G_{\phi}\right\vert \ll G_{t}$
whereas in case 2 $\left\vert G_{\phi}\right\vert $ can be larger than $G_{t}%
$. Thus we will study the consequences of the spin dependent BC for an
arbitrary value of $\left\vert G_{\phi}\right\vert /G_{t}$. In addition, in
case 1 we find $G_{\phi}<0$ but in case 2 the sign of $G_{\phi}$ depends on
the details of the barrier, thus we will consider both signs for $G_{\phi}$.

In equilibrium, we can use normal and anomalous quasiclassical Matsubara
Green's functions parametrized respectively as $\cos(\Lambda_{\sigma})$ and
$\sin(\Lambda_{\sigma})\exp(i\varphi_{\sigma})$ to describe the normal
excitations and the condensate of pairs (see e.g. \cite{RevueW}). The spatial
variations of the superconducting correlations in F are described by the
Usadel Eqs. $\partial Q_{\sigma}/\partial x=0$ and $\partial^{2}%
\Lambda_{\sigma}/\partial x^{2}=k_{\sigma}^{2}\mathrm{sgn}(\omega_{n}%
)\sin(\Lambda_{\sigma})/\xi_{F}^{2}+Q_{\sigma}^{2}$\textrm{cos}$(\Lambda
_{\sigma})/$\textrm{sin}$^{3}(\Lambda_{\sigma})$, with $\xi_{F}=(\hbar
D/E_{ex})^{1/2}$, $\omega_{n}=\left(  2n+1\right)  \pi k_{B}T$. Here,
$Q_{\sigma}=\sin^{2}(\Lambda_{\sigma})\partial\varphi_{\sigma}/\partial x$ is
the spectral current (constant with $x$) and $D$ the diffusion coefficient. We
introduced $k_{\sigma}=(2\left(  i\sigma\text{\textrm{sgn}}(\omega
_{n})+|\omega_{n}|/E_{ex}\right)  )^{1/2}$ for later use \cite{Golubov}.
Neglecting $G_{MR}$, $G_{\chi}$ and $G_{\xi}$ in (\ref{bc}) yields
\begin{align}
g_{F}\frac{\partial\Lambda_{\sigma}}{\partial x}  &  =iG_{\phi}\sigma
\sin(\Lambda_{\sigma})+G_{t}[\cos(\Lambda_{S})\sin(\Lambda_{\sigma
})\nonumber\\
&  -\sin(\Lambda_{S})\cos(\Lambda_{\sigma})\cos(\varphi_{\sigma}-\varphi_{S})]
\label{b2}%
\end{align}%
\begin{equation}
g_{F}\frac{\partial\varphi_{\sigma}}{\partial x}\sin(\Lambda_{\sigma}%
)=G_{t}\sin(\Lambda_{S})\sin(\varphi_{\sigma}-\varphi_{S}) \label{b1}%
\end{equation}
In Eqs. (\ref{b2}) and (\ref{b1}), we used rigid BC for S, i.e. $\Lambda
_{\sigma}=$ $\Lambda_{S}=\arctan[\Delta/\omega_{n}]$, with $\Delta$ the gap of S.

In the following, we consider the limit of a weak proximity effect in F, i.e.
$\Lambda_{\sigma}=\theta_{\sigma}$ for $\omega_{n}>0$ and $\Lambda_{\sigma
}=\pi-\theta_{\sigma}$ for $\omega_{n}<0$ with $|\theta_{\sigma}(x)|\ll1$. We
first study geometries with $Q_{\sigma}=0$, i.e. no supercurrent flows through
the device. In this case, the proximity effect in F can be probed through
measurements of the density of states $N(\varepsilon)=N_{0}(1-\sum_{\sigma
}\operatorname{Re}[\theta_{\sigma}^{2}(x)]/4)$ (with $\omega_{n}%
=-i\varepsilon+0^{+}$ and $\mathrm{sgn}(\omega_{n})=1$). The simplest case of
a single S/F interface with F at $x>0$ yields
\begin{equation}
\theta_{\sigma}^{SF}(x)=\frac{\gamma_{t}\sin(\Lambda_{S})}{\gamma
_{t}\left\vert \cos(\Lambda_{S})\right\vert +i\gamma_{\phi}\sigma
\mathrm{sgn}(\omega_{n})+k_{\sigma}}e^{-k_{\sigma}\frac{x}{\xi_{F}}}
\label{thetaSF}%
\end{equation}
with $\gamma_{t(\phi)}=G_{t(\phi)}\xi_{F}/g_{F}$. In the limit $\Delta\ll
E_{ex}$ where $k_{\sigma}=1+i\sigma$\textrm{sgn}$(\omega_{n})$, the weak
proximity effect hypothesis leading to (\ref{thetaSF}) is valid for any values
of $\gamma_{\phi}$ and $\varepsilon$ if $\gamma_{t}\ll1$. Since $k_{\sigma}$
has finite real and imaginary parts, $\theta_{\sigma}^{SF}(x)$ shows the
well-known exponentially damped sinusoidal oscillations with $d$. The
remarkable point in (\ref{thetaSF}) is that $\gamma_{\phi}$ \textit{shifts
these oscillations and modifies their amplitude} [see Fig. \ref{Graph1} which
shows the DOS following from (\ref{thetaSF})]. We also study the S/F/I
geometry, with F at $x\in\lbrack0,d]$ and the insulating layer I at $x>d$, for
later comparison with the experimental data of \cite{TakisN}. Using (\ref{b2})
for the S/F interface and $\partial\theta_{\sigma}/\partial x=0$ for F/I
yields:
\begin{equation}
\theta_{\sigma}^{SFI}(x)=\theta_{\sigma}^{d}\cosh\left(  (x-d)\frac{k_{\sigma
}}{\xi_{F}}\right)  \left[  \cosh\left(  k_{\sigma}\frac{d}{\xi_{F}}\right)
\right]  ^{-1} \label{thetaSFI}%
\end{equation}
with $\theta_{\sigma}^{d}=\gamma_{t}\sin(\Lambda_{S})/(\gamma_{t}|\cos
(\Lambda_{S})|+i\gamma_{\phi}\sigma\mathrm{sgn}(\omega_{n})+k_{\sigma}%
\tanh(k_{\sigma}d/\xi_{F}))$. In the limit $\Delta\ll E_{ex}$ and $d\geq
\xi_{F}$, the $\theta-$linearization leading to (\ref{thetaSFI}) is again
valid for any $\gamma_{\phi}$ and $\varepsilon$ if $\gamma_{t}\ll1$. From Fig.
\ref{Graph1}, $\gamma_{\phi}$ has qualitatively the same effect on
$\theta_{\sigma}^{SFI}(x)$ as on $\theta_{\sigma}^{SF}(x)$. More
quantitatively, for $d\gg\xi_{F}$ one has $\theta_{\sigma}^{SFI}%
(x=d)/\theta_{\sigma}^{SF}(x=d)=2$ \cite{Nremarl} and for lower values of $d$,
this ratio depends on $d$.

\begin{figure}[ptb]
\includegraphics[width=0.93\linewidth]{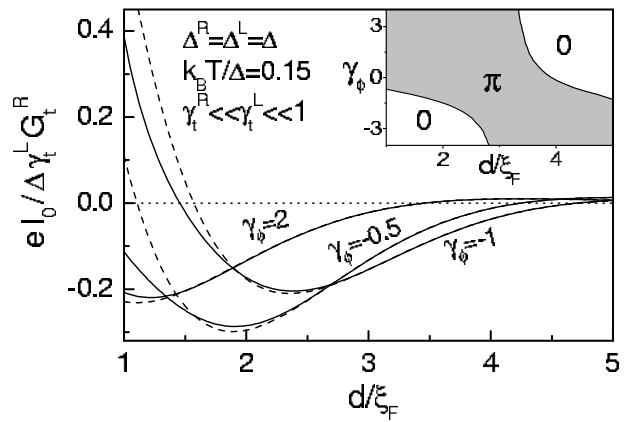}\caption{Critical current
$I_{0}$ of an asymetric S/F/S junction as a function of $d/\xi_{F}$,
calculated from Eq. (\ref{Qasym}) for $\gamma_{t}^{R}\ll\gamma_{t}^{L}\ll1$,
$\Delta^{L(R)}=\Delta\ll E_{ex}$ and $k_{B}T/\Delta=0.15$ (full lines). The
dashed lines show the large $d/\xi_{F}$ approximation of Eq. (\ref{asymLIM}).
The inset is a phase diagram indicating the equilibrium state of the junction
($0$ or $\pi$) depending on $\gamma_{\phi}$ and $d/\xi_{F}$.}%
\label{Graph2}%
\end{figure}

Another way to probe the superconducting proximity effect in F is to measure
the supercurrent through a S/F/S Josephson junction. We consider a junction
with F at $x\in\lbrack0,d]$ and a right(left) superconducting reservoir,
called $R$($L$) at a constant phase $(-)\varphi_{S}/2$.\ A supercurrent
$I_{S}=\pi g_{F}k_{B}T\sum\nolimits_{n\in\mathbb{Z},\sigma=\pm1}Q_{\sigma
}(\omega_{n})/2e$ flows through this device \cite{Golubov}. We focus on the
asymmetric limit $\gamma_{t}^{R}\ll\gamma_{t}^{L}$, which corresponds to the
experiment of \cite{TakisI}, and assume $\gamma_{\phi}^{R}=0$ \cite{comment}.
We allow $L$ and $R$ to have different superconducting gaps $\Delta^{R(L)}$,
so that $\Lambda_{\sigma}=\Lambda_{S}^{R(L)}$ in $R(L)$. Solving this problem
perturbatively with respect to the S/F/I case yields
\begin{equation}
Q_{\sigma}(\omega_{n})=\theta_{\sigma}^{d}\gamma_{t}^{R}\sin(\Lambda_{S}%
^{R})\sin(\varphi_{S})\left[  \xi_{F}\cosh\left(  k_{\sigma}\frac{d}{\xi_{F}%
}\right)  \right]  ^{-1} \label{Qasym}%
\end{equation}
where $\theta_{\sigma}^{d}$ corresponds to the expression given above with
$\Lambda_{S}=\Lambda_{S}^{L}$ and $\gamma_{t(\phi)}=\gamma_{t(\phi)}^{L}$. The
supercurrent has the form $I_{S}=I_{0}\sin(\varphi_{S})$ because most of the
phase drop occurs at $R$. In the limit $\Delta^{L}=\Delta^{R}=\Delta\ll
E_{ex}$, $\gamma_{t}^{L}\ll1$ and $d/\xi_{F}\gg1$, (\ref{Qasym}) yields
\begin{equation}
\frac{eI_{0}}{\gamma_{t}^{L}G_{t}^{R}\Delta}=\pi\tanh\left(  \frac{\Delta
}{2k_{B}T}\right)  \frac{\sin\left(  \frac{d}{\xi_{F}}+\lambda(\gamma_{\phi
}^{L})\right)  }{\left(  1+\left(  1+\gamma_{\phi}^{L}\right)  ^{2}\right)
^{\frac{1}{2}}}e^{-\frac{d}{\xi_{F}}} \label{asymLIM}%
\end{equation}
with $\lambda(\gamma_{\phi}^{L})=\arg[i-(1+\gamma_{\phi}^{L})]$. It is already
known that the state of the junction depends on $d$. Equation (\ref{asymLIM})
shows that $\gamma_{\phi}^{L}$ shifts the oscillations of the $I_{0}(d)$
curve. Thus, \textit{for a given value of }$d$, \textit{the state of the
junction can be }$0$ \textit{as well as }$\pi$, \textit{depending on}
$\gamma_{\phi}^{L}$. Fig. \ref{Graph2} shows that this effect still occurs
when one goes beyond the large $d/\xi_{F}$ approximation. Note that in the
limit $\Delta\ll E_{ex}$ and $\gamma_{t}^{L}\ll1$ used to obtain
(\ref{asymLIM}), it is not possible to find a temperature cross-over for the
sign of $I_{0}$ as observed in \cite{Ryazanov,SellierPRB}. However, we expect
to find such a temperature cross-over with a $0/\pi$ or $\pi/0$ transition,
depending on the value of $\gamma_{\phi}^{L}$, if the energy dependence of
$k_{\sigma}$ is taken into account \cite{later}. 

\begin{figure}[ptb]
\includegraphics[width=0.93\linewidth]{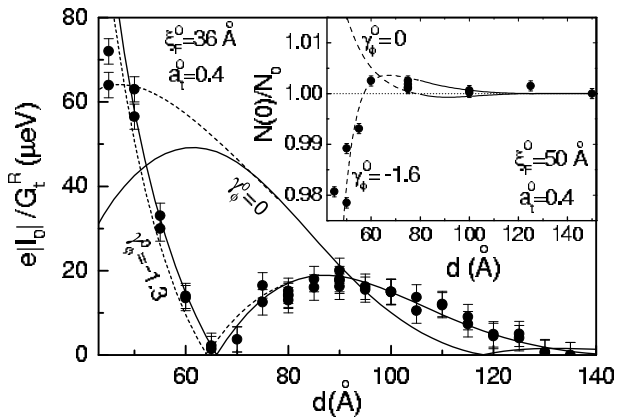}\caption{Critical current
measured by \cite{TakisI} through Nb/Pd$_{1-x}$Ni$_{x}$/Alox/Al/Nb junctions
as a function of the thickness $d$ of Pd$_{1-x}$Ni$_{x}$ (symbols). The lines
are theoretical curves calculated from Eq. (\ref{Qasym}) for $d>d_{0}$ (full
lines) and $d<d_{0}$ (dashed lines), with the fitting parameters $a_{t}=0.4$,
$\xi_{F}^{0}=36$~\textrm{\AA } and the experimentally determined parameters
$\Delta^{Nb}=1.35~$\textrm{meV}, $\Delta^{Al/Nb}=0.6~\mathrm{meV}$, $d_{0}%
=80$~\textrm{\AA } and $T=1.5~$\textrm{K}. The data are well fitted with
$\gamma_{\phi}^{0}=-1.3$. We also show the theory for $\gamma_{\phi}^{0}=0$.
Inset: DOS measured by \cite{TakisN} in Nb/Pd$_{1-x}$Ni$_{x}$/Alox/Al, as a
function of $d$. The full and dotted lines show the DOS at $x=d$ calculated
from the second-order generalization of (\ref{thetaSFI}) (see text), for
$d>d_{0}$ and $d<d_{0}$ respectively. We used $\xi_{F}^{0}=50$~\textrm{\AA }
and $\gamma_{\phi}^{0}=-1.6$ or $\gamma_{\phi}^{0}=0$, all the other
parameters being unchanged.}%
\label{Graph3}%
\end{figure}

To show the relevance of our approach, we compare our predictions with the
measurements of Refs. \cite{TakisN,TakisI}\textit{.} We first consider the
$\left\vert I_{0}\right\vert $ measured in an asymmetric S/F/S junction, i.e.
Nb/Pd$_{1-x}$Ni$_{x}$/Alox/Al/Nb with $x\sim0.1$ and $\gamma_{t}^{L}%
/\gamma_{t}^{R}\sim10^{5}$ \cite{TakisI}. We assume that the contacts have
$T_{n}\ll1$, which allows to use Eqs. (\ref{b2}-\ref{b1}). We will use the
experimentally determined values $\Delta^{Al/Nb}=0.6~$\textrm{meV} and
$\Delta^{Nb}=1.35$~\textrm{meV}$\ll E_{ex}$, which implies $k_{\sigma}%
\sim1+i\sigma$, and $T=1.5~$\textrm{K}. Samples with different thicknesses $d$
of PdNi were measured (see Fig. \ref{Graph3}). Interpreting these data
requires a careful analysis of the influence of $d$ on the different
parameters. We have $g_{F}=2e^{2}N_{0}DA$ and $\xi_{F}=\sqrt{\hbar D/E_{ex}}$,
with $D=v_{F}l/3$ and $A$ the conductors cross-section. Curie temperature
measurements show that the exchange field $E_{ex}$ increases linearly with $d$
\cite{TakisPhd}. In addition, we first assume that the mean free path $l$ is
constant with $d$, as confirmed by resistivity measurements for $d>d_{0}%
=80$~\textrm{\AA }. This allows to parametrize the problem with $\gamma
_{t}^{L}=a_{t}^{0}\xi_{F}^{0}/\sqrt{d_{0}d}$ and $\xi_{F}=\xi_{F}^{0}%
\sqrt{d_{0}/d}$ where $\xi_{F}^{0}$ and $a_{t}^{0}=G_{t}^{L}d_{0}/g_{F}$ are
constant with $d$. We also assume that $G_{\phi}^{L}$ is proportional to
$E_{ex}$ as found above in the rectangular barrier model for $E_{ex}\ll E_{F}%
$, so that we take $\gamma_{\phi}^{L}=\gamma_{\phi}^{0}\sqrt{d/d_{0}}$ with
$\gamma_{\phi}^{0}$ constant with $d$. We neglect $\gamma_{\phi}^{R}$ due to
the existence of a strong insulating barrier at R \cite{comment}. The absolute
amplitude of $E_{ex}$ was not determined exactly, so that $\xi_{F}^{0}$ can be
considered as a fitting parameter as well as $a_{t}^{0}$ and $\gamma_{\phi
}^{0}$. This makes in total three fitting parameters but we expect to find for
$a_{t}^{0}$ a value close to the value $0.2$ found from minigap measurements
in Nb/Pd \cite{TakisPhd}. We have calculated $\left\vert I_{0}\right\vert $ by
summing (\ref{Qasym}) on energy and spin. It is not possible to account for
the data with $\gamma_{\phi}^{0}=0$. On the contrary, a good agreement with
the experiment is obtained by using $a_{t}^{0}=0.4$, $\xi_{F}^{0}%
=36~$\textrm{\AA } and $\gamma_{\phi}^{0}=-1.3$\textrm{\ }(full lines in Fig.
\ref{Graph3}) \cite{KontosFitC,buzdinFitC}. We have checked that this choice
of parameters fulfills the hypothesis $\left\vert \theta_{\sigma
}(x)\right\vert \ll1$ made in our calculations. Remarkably, for $d\sim d_{0}$
in Fig. \ref{Graph3}, the theory for $\gamma_{\phi}^{0}=-1.3$ gives $I_{0}<0$
in agreement with subsequent experiments \cite{Bauer}, whereas it gives
$I_{0}>0$ for $\gamma_{\phi}^{0}=0$ if one keeps the same orders of magnitude
for $a_{t}^{0}$ and $\xi_{F}^{0}$. For $d<d_{0}$, $l$ is linear with $d$,
which we have taken into account by using $\xi_{F}=\xi_{F}^{0}$, $\gamma
_{\phi}^{L}=\gamma_{\phi}^{0}$, and $\gamma_{t}^{L}=a_{t}^{0}\xi_{F}^{0}/d$,
with the same values of $a_{t}^{0}$, $\gamma_{\phi}^{0}$ and $\xi_{F}^{0}$ as
previously (dashed lines in Fig. \ref{Graph3}). This approach gives a
surprising agreement with the data, which seems to indicate that the Usadel
description still works for $d<d_{0}$ although $l$ is linear with $d$
\cite{Maleck}. Kontos \textit{et al.} have also performed DOS measurements in
Nb/Pd$_{1-x}$Ni$_{x}$/Alox/Al \cite{TakisN}, prior to the $I_{0}$ measurements
\cite{KontosFitC}. We have assumed again that $E_{ex}$ was linear with $d$ in these
measurements, to try to interpret the $N(0)=f(d)$ curve with the same fitting
procedure as for $I_{0}$. We have generalized Eq. (\ref{thetaSFI}) to second
order in $\theta_{\sigma}$ because the values of $d/\xi_{F}$ are slightly
lower than for the $I_{0}$ measurements. Again it is impossible to interpret
the data with $\gamma_{\phi}^{0}=0$. We obtain a satisfactory fit by choosing
$\xi_{F}^{0}=50~$\textrm{\AA } and $\gamma_{\phi}^{0}=-1.6$, all the other
parameters used being the same as in the previous case. Finding a $\xi_{F}%
^{0}$ higher than for the $I_{0}$ data is in agreement with the fact that the
samples used for measuring the DOS were realized with a lower concentration
$x$ of Ni.

In summary, we have studied the effect of spin-dependent boundary conditions
on the superconducting proximity effect in a diffusive ferromagnet bounded by
tunneling interfaces. We have shown that the phase-shifting conductances
$G_{\phi}$, describing the spin-activity of the interfaces in this context,
strongly affect the behavior of the system and allow a consistent microscopic
explanation of the DOS and supercurrent data of \cite{TakisN,TakisI}\textit{.}
This suggests that such effects will have to be considered in any future work
on S/F hybrid circuits. In the context of spintronics, this approach might
also provide a way to characterize spin-active interfaces.

We thank T. Kontos for raising the question which led us to perform this study
and for providing us with the experimental data. We thank C. Bruder, T.T.
Heikkil\"{a} and D. Huertas-Hernando for discussions. This work was financed
by the Swiss NSF and the NCCR Nanoscience.


\begin{thebibliography}{99}
\bibitem {comment}With a high (rectangular) barrier, one has $G_{t}%
\ll\left\vert G_{\phi}\right\vert $, but if the barrier is too high, one can
check that $I_{0}$ is unaffected by $G_{\phi}^{R}$ because $G_{\phi}^{R}\ll
G_{t}^{L},G_{\phi}^{L}$.
\bibitem {buzdinFitC}Another interpretation of \cite{TakisI} was proposed in
\cite{buzdinFit} for $\gamma_{\phi}=0$, but with values of $\Delta_{L}%
\Delta_{R}$ and $T$ in disagreement with the experimentally determined
parameters and continuous BC $\Lambda_{\sigma}=\Lambda_{S}^{L}$ for $x=0$.
\end{thebibliography}
\end{document}